*Original Research Article*

# Charged Perfect Fluid Distribution for Cosmological Universe Interacting With Massive Scalar Field in Brans-Dicke Theory.

**Kangujam Priyokumar Singh**
*Department of Mathematical Sciences, Bodoland University, Kokrajhar , BTC, Assam (INDIA).*
*E-mail:- pk_mathematics@yahoo.co*.in ; *pk_mathematicsbu@hotmail.com*

**Abstract**

Considering a spherically-symmetric non-static cosmological flat model of Robertson-Walker universe we have investigated the problem of perfect fluid distribution interacting with the gravitational field in presence of massive scalar field and electromagnetic field in B-D theory. Exact solutions have been obtained by using a general approach of solving the partial differential equations and it has been observed that the electromagnetic field cannot survive for the cosmological flat model due to the influence caused by the presence of massive scalar field.

## 1. Introduction

Apart from the presence of charge, which is responsible for avoiding the gravitational collapse of spherically symmetric fluid distributions of the matter to a point due to the Coulombian repulsive forces along with the pressure gradient? Many of the authors charged the well-known uncharged perfect fluid solutions, also they have been investigated the problems of scalar field with and without a mass parameter satisfying the Klein-Gordan equation. Rao and Roy [1,2] have investigated the problem of massive scalar field interacting with or without the presence of electromagnetic field and have found that massive scalar field cannot survive for the axially symmetric and further it has been found that massive scalar field cannot interact with the electromagnetic field for the axially symmetric metric. Candelas [3] has proposed that by considering a massive scalar field in general theory of relativity the emergence of singularities can be avoided. Das and Agrawal [4] investigated the interactions of massive scalar and electromagnetic field in Friedmann universe. Rao et al.[5] have obtained physically meaningful solutions corresponding to Robertson-Walker metric by taking the massive scalar field as source and observed that the mass parameter in the Klein-Gordon equation corresponding to the mass of the gravitation. Mohanty and Panigrahi [6] have presented a technique which enables one to generate a class of solutions to Einstein-Maxwell-Massless scalar field equations with a self gravitating stiff perfect fluid as source of gravitation. Tarachand and Ibotombi [7] have investigated about the problem of non existence of interactions between axially symmetric massive scalar field and perfect fluid distributions by considering the axially symmetric Einstein-Rosen metric. They observed that there cannot exist any solution for the coupled massive scalar field and perfect fluid distribution for the metric. The problem reduces to the finding of interactions between zero-mass scalar field and stiff fluid. Ibotombi and Gokulchandra [8] studied the problem of cosmological massive scalar field interacting with viscous fluid by considering the source density in the wave equation subject to certain physical conditions. Some other researchers who have the contribution in this area are Gupta and Kumar [9], Bijalwan and Gupta [10], Gupta and Maurya [11, 12, 13] etc.

In the present paper we have investigated the problem of cosmological charged perfect fluid distributions interacting with massive scalar field in B-D theory by considering a spherically-symmetric non-static cosmological flat model of Robertson-Walker universe. We have observed that the electromagnetic field cannot service for cosmological flat model due to the influence caused by the presence of massive scalar field.

In section 2 we have presented the formulation of the problem and their solutions. In section 3 we have presented the physical interpretation of the solution obtained.

## 2. Formulation of the Problem and their Solutions

Nordtvedl's field equations in Dicke's conformally transformed units are given by

$$G^i_j \equiv R^i_j - \frac{1}{2}\delta^i_j R = -8\pi T^i_j - \frac{2\omega+3}{2\phi^2}\left[\phi^{,i}\phi_{,j} - \frac{1}{2}\delta^i_j \phi^{,m}\phi_{,m}\right] \tag{1}$$

and the wave equation for the scalar field is

$$\Box(\ell n\phi) \equiv (\ell n\phi)_{;m}^{;m} = \frac{8\pi}{2\omega+3}T, \tag{2}$$

The energy-momentum tensor for charged perfect fluid distribution interacting with massive scalar field is given by

$$T_j^i \equiv Z_j^i + E_j^i + S_j^i$$

$$= \left[(p+\rho)u^i u_j - p\delta_j^i + \frac{1}{4\pi}\left\{-F^{i\alpha}F_{j\alpha} + \frac{1}{4}\delta_j^i F_{\alpha\ell} F^{\alpha\ell}\right\}\right.$$

$$\left. + \frac{1}{4\pi}\left\{V^{,i} V_{,j} - \frac{1}{2}\delta_j^i \left(V^{,m} V_{,m} - M^2 V^2\right)\right\}\right] \tag{3}$$

together with

$$u^i u_i = 1, \tag{4}$$

where p is the isotropic pressure, $\rho$ the fluid density, V the scalar field, $u_i$ the four velocity vector, $F_{ij}$ the electromagnetic field tensors $Z_j^i$, $E_j^i$ and $S_j^i$ denote the energy-momentum tensors for perfect fluid, electromagnetic field and massive scalar field respectively and are given by

$$Z_j^i = (p+\rho)u^i u_j - p\delta_j^i, \tag{5}$$

$$E_j^i = \frac{1}{4\pi}\left\{-F^{i\alpha} F_{j\alpha} + \frac{1}{4}\delta_j^i F_{\alpha\ell} F^{\alpha\ell}\right\} \tag{6}$$

and

$$S_j^i = \frac{1}{4\pi}\left\{V^{,i} V_{,j} - \frac{1}{2}\delta_j^i \left(V^{,m} V_{,m} - u^2 V^2\right)\right\} \tag{7}$$

Using the co-moving co-ordinate system, we obtain

$$u^1 = u^2 = u^3 = 0 \quad \text{and} \quad u^4 = 1 \tag{8}$$

The electromagnetic field equations are given by

$$F_{;j}^{ij} = -J^i \tag{9}$$

and

$$F_{[ij,k]} = 0, \tag{10}$$

where $J^i$ is the current 4-vector and in general, is the sum of the convention current and conduction current i.e.,

$$J^i = \epsilon_0 u^i - \sigma_0 u^v F_v^i, \tag{11}$$

where $\epsilon_0$ is the rest charge density and $\sigma_0$ is the conductivity coefficient.

The scalar potential V satisfies the Klein-Gordon equation

$$g^{ij} V_{;ij} + M^2 V = \sigma(r,t), \tag{12}$$

where $\sigma(r, t)$ is the source density of the scalar potential and M is related to the mass of zero-spin particle by

$$M = \frac{m}{\hbar}, \quad \text{where} \quad \hbar = \frac{h}{8\pi},$$

h being the Planck's constant.

The metric taken for the present problem is

$$ds^2 = dt^2 - R^2(t)\left[\frac{dr^2}{1-Kr^2} + r^2 d\theta^2 + r^2 \sin^2\theta d\phi^2\right], \tag{13}$$

where t is the cosmic time, R(t) the scale factor of the universe and K the curvature index which takes up the values +1, 0, -1. Here a comma or semi colon followed by a subscript denotes partial or covariant differentiation respectively. A dot and a dash over a letter denote partial differentiation with respect to time t and radial distance r respectively.

The surviving field equations for the metric (13) are

$$G_1^1 \equiv \frac{2\ddot{R}}{R} + \frac{\dot{R}^2}{R^2} + \frac{K}{R^2}$$

$$= -8\pi p + 8\pi E_1^1 - \frac{1-Kr^2}{R^2} V'^2 - (\dot{V}^2 - M^2 V^2) -$$

$$- \frac{2\omega+3}{4}\left\{\frac{1-Kr^2}{R^2}\left(\frac{\phi'}{\phi}\right)^2 + \left(\frac{\dot{\phi}}{\phi}\right)^2\right\}, \tag{14}$$

$$G_2^2 \equiv \frac{2\ddot{R}}{R} + \frac{\dot{R}^2}{R^2} + \frac{K}{R^2}$$

$$= -8\pi p + 8\pi E_2^2 + \frac{1-Kr^2}{R^2} V'^2 - (\dot{V}^2 - M^2 V^2) +$$

$$+ \frac{2\omega+3}{4}\left\{\frac{1-Kr^2}{R^2}\left(\frac{\phi'}{\phi}\right)^2 - \left(\frac{\dot{\phi}}{\phi}\right)^2\right\}, \tag{15}$$

$$G_2^2 \equiv G_2^2, \tag{16}$$

$$G_4^4 \equiv 3\left[\frac{\dot{R}^2}{R^2} + \frac{K}{R^2}\right]$$

$$= 8\pi\rho + 8\pi E_4^4 + \frac{1-Kr^2}{R^2}V'^2 + (\dot{V}^2 + M^2V^2) +$$

$$+ \frac{2\omega+3}{4}\left\{\frac{1-Kr^2}{R^2}\left(\frac{\phi'}{\phi}\right)^2 + \left(\frac{\dot{\phi}}{\phi}\right)^2\right\}, \tag{17}$$

$$G_4^1 \equiv \sigma = 4\pi E_4^1 + V'\dot{V} + \frac{2\omega+3}{4}\left(\frac{\phi'}{\phi}\right)\left(\frac{\dot{\phi}}{\phi}\right), \tag{18}$$

$$G_2^1 \equiv 0 = \frac{1}{R^4 r^2 \sin^2\theta} F_{13} F_{23} - F_{14} F_{24}, \tag{19}$$

$$G_3^1 \equiv 0 = \frac{1}{R^2 r^2} F_{12} F_{23} + F_{14} F_{34}, \tag{20}$$

$$G_3^2 \equiv 0 = -\frac{1-Kr^2}{R^2} F_{12} F_{13} + F_{24} F_{34}, \tag{21}$$

$$G_4^2 \equiv 0 = -(1-Kr^2)F_{12}F_{14} + \frac{1}{r^2\sin^2\theta}F_{23}F_{34} \tag{22}$$

and

$$G_4^3 \equiv 0 = (1-Kr^2)F_{13}F_{14} + \frac{1}{r^2}F_{23}F_{24}. \tag{23}$$

The wave equation (2) reduces to

$$(3+2\omega)\left[-\frac{1-Kr^2}{R^2}(\ell n\phi)'' - \frac{2-3Kr^2}{R^2 r}(\ell n\phi)' + 3\frac{\dot{R}}{R}(\ell n\phi)^{\cdot} + (\ell n\phi)^{\cdot\cdot}\right]$$

$$= 8\pi(\rho - 3p) + 2\left\{\frac{1-Kr^2}{R^2}V'^2 - \dot{V}^2 + 2M^2V^2\right\}. \tag{24}$$

Also from equation (12) we have

$$-\frac{1-Kr^2}{R^2}V'' - \frac{2-3Kr^2}{R^2 r}V' + 3\frac{\dot{R}}{R}\dot{V} + \ddot{V} + M^2 V = \sigma(r,t). \tag{25}$$

Again, from equation (9) we have the following four equations

$$\frac{\partial F^{14}}{\partial t} + \frac{3\dot{R}}{R}F^{14} = \sigma F_4^1, \tag{26}$$

$$\frac{\partial F^{14}}{\partial r} + \left(\frac{Kr}{1-Kr^2} + \frac{2}{r}\right)F^{14} = \epsilon_0, \tag{27}$$

$$\frac{\partial F^{23}}{\partial \theta} + F^{23}\cot\theta = 0 \tag{28}$$

and

$$\frac{\partial F_{14}}{\partial \phi} = 0 \tag{29}$$

Also from equation (10) we have

$$\frac{\partial F_{14}}{\partial \theta} = \frac{\partial F_{14}}{\partial \phi} = 0 \tag{30}$$

and

$$\frac{\partial F_{23}}{\partial r} = \frac{\partial F_{23}}{\partial t} = 0. \tag{31}$$

From equations (19) to (23) we obtain the following three possible cases :

(i) $F_{12} = F_{13} = F_{34} = F_{24} = 0$ at least one of $F_{14}$, $F_{23}$ being non-zero.

(ii) $F_{12} = F_{14} = F_{34} = F_{23} = 0$ at least one of $F_{24}$, $F_{13}$ being non-zero.

(iii) $F_{14} = F_{24} = F_{13} = F_{23} = 0$ at least one of $F_{12}$, $F_{34}$ being non-zero.

Hence, the electromagnetic field is non-null and consists of an electric and / or magnetic field both of which are in the direction of the same space axis. Without loss of generality, we may consider the case (i) in which also the component $F_{14} \neq 0$, $F_{23} = 0$, which is directed in the direction of x-axis.

Using the above assumption in equation (6), we obtain

$$E_1^1 = -E_2^2 = -E_3^3 = E_4^4 = \frac{1}{8\pi} \cdot \frac{1-Kr^2}{R^2}(F_{14})^2 \tag{32}$$

and
$$E^i_j = 0, \quad (i \neq j, \; i, j = 1, 2, 3, 4). \tag{33}$$

Making use of equation (32) the field equations (14), (15) and (17) respectively become

$$\frac{2\ddot{R}}{R} + \frac{\dot{R}^2}{R^2} + \frac{K}{R^2} = -8\pi p + \frac{1-Kr^2}{R^2}(F_{14})^2 - \frac{1-Kr^2}{R^2}V'^2 - (\dot{V}^2 - M^2V^2) -$$
$$- \frac{2\omega + 3}{3}\left\{\frac{1-Kr^2}{R^2}\left(\frac{\phi'}{\phi}\right)^2 + \left(\frac{\dot{\phi}}{\phi}\right)^2\right\}, \tag{34}$$

$$2\frac{\ddot{R}}{R} + \frac{\dot{R}^2}{R^2} + \frac{K}{R^2} = -8\pi p - \frac{1-Kr^2}{R^2}(F_{14})^2 + \frac{1-Kr^2}{R^2}V'^2 - (\dot{V}^2 - M^2V^2) +$$
$$+ \frac{2\omega + 3}{4}\left\{\frac{1-Kr^2}{R^2}\left(\frac{\phi'}{\phi}\right)^2 - \left(\frac{\dot{\phi}}{\phi}\right)^2\right\} \tag{35}$$

and

$$\frac{3\dot{R}^2}{R^2} + \frac{3K}{R^2} = 8\pi\rho + \frac{1-Kr^2}{R^2}(F_{14})^2 + \frac{1-Kr^2}{R^2}V'^2 - (\dot{V}^2 - M^2V^2) +$$
$$+ \frac{2\omega + 3}{4}\left\{\frac{1-Kr^2}{R^2}\left(\frac{\phi'}{\phi}\right)^2 + \left(\frac{\dot{\phi}}{\phi}\right)^2\right\}. \tag{36}$$

Also using equation (33) in equation (18), we obtain

$$V'\dot{V} + \frac{2\omega + 3}{4}\left(\frac{\phi'}{\phi}\right)\left(\frac{\dot{\phi}}{\phi}\right) = 0. \tag{37}$$

Solving equation (37), we obtain

$$V = \frac{\sqrt{-(2\omega + 3)}}{2}\log\phi + A_1. \tag{38}$$

where $A_1$ is an arbitrary constant.

From equations (34), (35) and (38), we obtain

$$F_{14} = 0 \tag{39}$$

From equations (35) and (39), we obtain

$$8\pi p = M^2V^2 - \left(\frac{2\ddot{R}}{R} + \frac{\dot{R}^2}{R^2} + \frac{K}{R^2}\right). \tag{40}$$

From equations (36) and (39), we obtain

$$8\pi\rho = \frac{3\dot{R}^2}{R^2} + \frac{3K}{R^2} - M^2 V^2. \tag{41}$$

For flat model of the universe, i.e., $K = 0$, we obtain from equation (24)

$$-\frac{1}{R^2}(\ell n\phi)'' + \frac{2}{R^2 r}(\ell n\phi)' + 3\frac{\dot{R}}{R}(\ell n\phi)^{\cdot} + (\ell n\phi)^{\cdot\cdot} = \frac{8\pi}{3+2\omega}(\rho - 3p) +$$

$$+ \frac{2}{3+2\omega}\left\{\frac{1}{R^2}V'^2 - \dot{V}^2 + 2M^2 V^2\right\} \tag{42}$$

Using the well known Hubble's principle

$$\frac{\dot{R}}{R} = H, \tag{43}$$

where H is the Hubble's constant.

Using equation (43) in equation (42), we obtain

$$\frac{3}{2}\left(\frac{\phi'}{\phi}\right)^2 - \frac{\phi''}{\phi} - \frac{2}{r}\frac{\phi'}{\phi} = R^2\left[\frac{12H^2}{3+2\omega} + \frac{3}{2}\left(\frac{\dot{\phi}}{\phi}\right)^2 - 3H\left(\frac{\dot{\phi}}{\phi}\right) - \frac{\ddot{\phi}}{\phi}\right] \tag{44}$$

From equation (44), we obtain

$$\phi = e^{\alpha t} \cdot \frac{r^2}{(A_2 - Br)^2} \tag{45}$$

with relation on constants

$$24H^2 + (\alpha^2 - 6H\alpha)(2\omega + 3) = 0, \tag{46}$$

where $\alpha$, $A_2$ and B are arbitrary constants.

From equations (38) and (45), we get

$$V = \frac{\sqrt{-(2\omega + 3)}}{2}\log\left\{\frac{e^{\alpha t} \cdot r^2}{(A_2 - Br)^2}\right\} + A_1 \tag{47}$$

For the flat model of the universe, the source density is given by

$$\sigma(r, t) = \sqrt{-(2\omega + 3)}\left\{\frac{3}{2}\alpha H - \frac{A_2^2 e^{-2Ht}}{C_1^2 r^2(A_2 - Br)}\right\} +$$

$$+ M^2\left[\frac{\sqrt{-(2\omega + 3)}}{2}\log\left\{\frac{e^{\alpha t} \cdot r^2}{(A_2 - B_2)}\right\} + A_1\right], \tag{48}$$

where $C_1$ is an arbitrary constant.

The isotropic pressure p and the fluid density $\rho$ are given by

$$p = \frac{1}{8\pi}\left[M^2 V^2 - 3H^2\right] \tag{49}$$

and

$$\rho = \frac{1}{8\pi}\left[3H^2 - M^2 V^2\right], \tag{50}$$

where V is given in equation (47).

From equation (27), we obtain

$$\epsilon_0 = 0. \tag{51}$$

From equation (11), we obtain

$$J^i = 0 \; ; \; i = 1, 2, 3, 4. \tag{52}$$

### 3. Physical Interpretation of the Solutions Obtained

It is evident from equation (39) that the electromagnetic field when interacting with massive scalar field and Brans-Dicke Scalar field $\phi$ in the presence of perfect fluid does not survive. From equation (45), it is found that the B-D scalar $\phi$ is an exponentially increasing function of time provided $24H^2 + (\alpha^2 - 6\alpha H)(2\omega + 3) = 0$ and $A_2 = 0$, and it is a quadritically increasing function of radial co-ordinate 'r' when B = 0 at a certain instant. Keeping the radial co-ordinate, r as constant, the Brans-Dicke scalar $\phi$ tends to infinity as cosmic time t tends to infinity. From equation (47), we observed that the massive scalar field V is physically realistic provided the coupling constant $\omega < -3/2$. It is further observed that the massive scalar field V becomes a linear function of time t when $A_2 = 0$. The source density '$\sigma$' of the scalar potential reduces to a constant quantity when both r and t tend to infinity in the case of M = 0, i.e., zero-rest-mass scalar field in B-D Theory. For realistic solution $\rho \geq 0$ when $\rho > 0$, the isotropic pressure p is found to be negative. It happens due to the interaction caused by the presence of massive scalar field V and the fluid acquires a repulsive character in nature. The electromagnetic field gives no contribution in yielding the isotropic pressure p to be positive. The matter becomes electrically neutral as the charged density $\epsilon_0$ and the current density $J_i$ vanish. From equations (49) and (50) we found that $p + \rho = 0$ which shows that the universe does not obey the fluid energy condition $p + \rho > 0$.

.